# Sub nanosecond coherent optical manipulation of a single aromatic molecule at cryogenic temperature.

Quentin Deplano[1,2], Philippe Tamarat[1,2], Brahim Lounis[1,2], and Jean-Baptiste Trebbia[1,2]

*1- Univ Bordeaux, LP2N, F- 33405 Talence, France*
*2- Institut d'Optique & CNRS, LP2N, F-33405 Talence, France*

Corresponding author : Jean-Baptiste Trebbia
email: jean-baptiste.trebbia@u-bordeaux.fr

## Abstract

Single molecules trapped in the solid state at liquid helium temperatures are promising quantum emitters for the development of quantum technologies, owing to their remarkable photostability and their lifetime-limited optical coherence time of the order of ten nanoseconds. The coherent preparation of their electronic state requires resonant excitation with a Rabi period much shorter than their optical coherence time. Sculpting the optical excitation with sharp edges and a high on-off intensity ratio ($\sim 3\times 10^5$) from a single-frequency laser beam, we demonstrate sub-nanosecond drive of a single dibenzanthanthrene molecule embedded in a naphthalene matrix at 3.2 K, over more than 17 Rabi periods. With pulses tailored for a half-Rabi period, the electronic excited state is prepared with fidelity as high as 0.97. Using single-molecule Ramsey spectroscopy, we prove up to 5 K that the optical coherence lifetime remains at its fundamental upper limit set by twice the excited state lifetime, making single molecules suitable for quantum bit manipulations under standard cryogen-free cooling technologies.

## I. INTRODUCTION

The coherent control of photon-matter interfaces is an interesting approach for the realization of quantum information networks. A large variety of optically addressable qubit platforms have been developed on the basis of ultra cold atoms [1,2], Silicon-Vacancy defects in diamond [3,4] or semi-conducting quantum dots [5-7]. Among those systems, organic molecules trapped in condensed matter at liquid helium temperatures are particularly promising candidates [8]. They behave as two-level systems with excellent photostability [9] and display zero-phonon transition lines (ZPLs) with coherence times $T_2$ limited by the excited-state lifetime $T_1$ ($T_2 = 2T_1$) [10]. These extended coherence times, up to tens of ns [11], make them suitable for quantum interference experiments [12-14]. Moreover, they have proven their effectiveness for the generation of triggered single photons [15,16], with potential applications for quantum key distribution [17,18]. The controlled interaction of single molecules with photonic structures [19-23] also leads to enhanced interaction efficiencies with light, offering interesting prospects for integrated quantum photonics. On the other hand, the possibility of fast, coherent optical manipulation of the molecular electronic states has been poorly investigated. To effectively invert the populations of the ground and excited states, or to create a superposition of these states with maximum coherence, we would need tailored resonant pulses associated with a quarter or half of a Rabi period, respectively. Near-unity fidelity in the state preparation will require that the damping of the transition dipole be negligible during the preparation time for the molecular state. It is thus crucial for the Rabi period to be much shorter than $T_2$.

Up to now, optical nutation of single molecules have been monitored under resonant pulsed excitation [24,25], in a similar way to what was performed with atoms [26] or solid-state quantum emitters such as Silicon-Vacancy defect centers in diamond [27,28]. Yet, their modest contrast would not allow to prepare the excited state with a fidelity better than ~ 0.7. In this letter, we report optical pulse shaping with sharp edges and a high on-off intensity ratio from a resonant single-frequency laser, and demonstrate sub-ns optical nutation in the fluorescence of an isolated single molecule. Negligible damping of the transition dipole over the first period shows the possibility to prepare the molecular excited state with a probability 0.97 after a π-pulse excitation. Moreover, using single-molecule optical Ramsey spectroscopy [29], we directly measure the dephasing time $T_2$ and prove up to 5 K that it remains at its fundamental upper limit $2T_1$, making molecular electronic states suitable qubits for quantum logic gates operating under standard cryogen-free cooling technologies.

## II. EXPERIMENTAL SETUP

The sample used in this study is prepared by heating a piece of naphthalene doped with dibenzanthanthrene (DBATT) (Fig. 1a) above the melting temperature and pressing it between two coverslips to obtain a thin layer with a thickness of a few μm. The sample is then cooled down to 3.2 K in a cryogen-free cryostat. A single-frequency tunable dye laser (linewidth ~1 MHz, wavelength ~618 nm) is used to excite the purely electronic transition of the molecules placed at the diffraction-limited focal spot (transverse FWHM ~ 500 nm) of a microscope objective with numerical aperture 0.95. The red-shifted fluorescence of the excited molecules is sent through a band-pass filter (bandwidth 100 nm centered on 700 nm) onto a single-photon avalanche photodiode with temporal resolution ~50 ps at half maximum. The response function of the avalanche photodiode to femtosecond pulses is displayed in Fig. S1.

The DBATT single molecule chosen for this study has its transition dipole moment aligned with the linear polarization of the laser beam, in order to keep the saturation intensity of the ZPL extremely low. This allows fast optical drive of its Bloch vector with modest excitation powers that will hardly heat the sample and generate thermal dephasing. Figure 1(b) shows its fluorescence excitation spectrum in the low saturation regime at 3.2 K. The optical resonance displays a Lorentzian profile with a FWHM ZPL width $\Delta \nu_0 = (22.3 \pm 0.07)$ MHz, in agreement with the previous values reported in the literature for this host-guest system [30]. From this linewidth, one can estimate the dephasing time of the optical transition $T_2 = (\pi \Delta \nu_0)^{-1} = (14.1 \pm 0.05)$ ns. The evolution of the ZPL width with increasing excitation intensities $I$ (Fig. 1c) is consistent with the saturation law $\Delta \nu(I) = \Delta \nu_0 \sqrt{1 + I/I_{sat}}$, which provides the saturation intensity $I_{sat} = 0.32$ W cm$^{-2}$ of this molecule. The absence of deviation from the saturation law up to the strongest excitation intensities $\sim 10^4\ I_{sat}$ is a signature of negligible dephasing due to laser-induced local heating processes, such as non-radiative vibrational relaxation that accompanies fluorescence. After switching off the laser beam with an electro-optic modulator (EOM), the fluorescence signal of the molecule decays exponentially, as shown in the inset of Fig. 1c). This decay provides a measure of the electronic excited state lifetime $T_1 = (7.2 \pm 0.3)$ ns, which is consistent with $T_2 \cong 2T_1$.

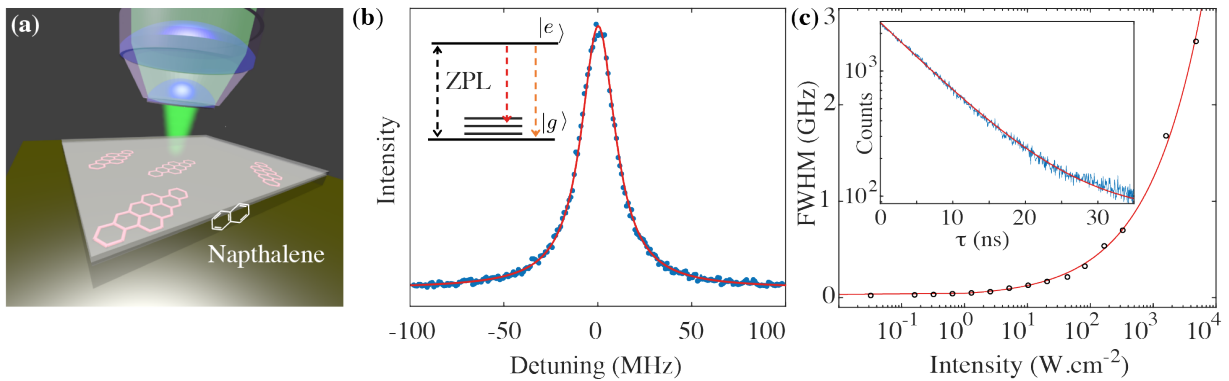

Figure 1 : **a)** Illustration of the single-molecule microscopy setup. The naphthalene matrix doped with DBATT is crystallized on a glass coverslip. A microscope objective focuses the laser to excite the molecules and collect its red-shifted fluorescence. **b)** Fluorescence excitation spectrum of a single DBATT molecule at 3.2 K, recorded in the low saturation regime ($I/I_{sat} \sim 0.2$). Its signal-to-background ratio is greater than 200. The spectrum is averaged over 40 spectra (5 ms per bin) that have been centered to eliminate residual broadening by low-frequency spectral diffusion. The inset is a diagram of the ground and first excited electronic levels of the molecule, as well as the vibrational levels involved in the fluorescence transitions. **c)** Evolution of the homogeneous spectral broadening with increasing excitation intensities for the same molecule as in **b**). The red curve is a fit with the saturation law. The inset shows the fluorescence decay subsequent to a resonant excitation shaped with a single EOM (30 ns duration, start at time 0, $I/I_{sat} \sim 5$), for the same molecule. Fitting with an exponential decay (red curve) is made several nanoseconds after the falling edge of the excitation to avoid deconvolution with the EOM falling edge. The exponential fit provides the excited-state lifetime $T_1 = (7.2 \pm 0.3)$ ns.

## III. OPTICAL NUTATION

Coherent optical manipulation of a single molecule with a high fidelity requires strong excitation intensities with abrupt on and off switches, as well as a high on-off intensity ratio to avoid optical saturation of the molecule during the off periods. The experimental setup built to obtain fast, sub-ns optical nutation is depicted in Fig. 2a. As a first step, the resonant optical excitation is prepared by shaping the laser beam with just an EOM driven by a square-wave RF signal, leading to an on-off intensity ratio $\kappa \sim 8 \times 10^3$ (See Fig. S2 for details on the EOM operation). Figure 2b shows the fluorescence time trace of the molecule subjected to resonant excitation for a duration of 30 ns with strong intensity $I \sim 6100\ I_{sat}$. This evolution is displayed after background subtraction (See Figs S3 and S4) and deconvolution from the response function of the avalanche photodiode, as described in Fig. S5. Rabi oscillations clearly show up with a period of 850 ps and are damped throughout the illumination slot as a result of optical coherence relaxation. At the end of the excitation, the molecule reaches its steady state characterized by a saturated population 0.5 of the excited state, which allows rescaling the fluorescence signal into excited-state population. Such normalization allows reproducing the experimental time trace with the computed evolution of the excited-state population (red curve), derived from the semi-classical Bloch equations of a two-level system under resonant excitation. Before the laser excitation is switched on ($t < 0$ in Fig. 2b), this population is kept at ~ 0.2 as a result of the residual laser intensity during the off-periods of the EOM. This artifact degrades the contrast of the Rabi oscillations and would limit to ~ 0.9 the probability to prepare the molecular excited state after pulsed excitation for half a Rabi period ($\pi$-pulse). The fidelity in such preparation is simulated in Fig. 2c as a function of the on-off intensity ratio $\kappa$, for an intensity $I = 6100\ I_{\text{sat}}$. It shows that the fidelity is expected to approach unity (~ 0.97) for values of $\kappa$ above $10^5$.

In order to properly initialize the molecule in the ground state before switching on the EOM, the on-off intensity ratio is enhanced by adding a fast acousto-optic modulator (AOM), which commutates the laser beam in its first diffraction order before injection into the EOM fiber (Fig.2a). The electronic signals commanding these modulators are delivered by an arbitrary-wave generator and finely shaped to sculpt the excitation slots with flat-top profiles and with an on-off laser intensity ratio $\kappa \sim 3\times10^5$ (See Fig. S6 for details on the pulse profile). With such pulse shaping, optical nutation of the molecule can be driven at high frequencies and with enhanced visibility, as demonstrated in Fig. 2d for $I/I_{\text{sat}} = 6100$. At least 17 Rabi oscillations can be distinguished in Fig. 2d, which to our knowledge constitutes the largest number of optical Rabi oscillations observed to date with a single solid-state emitter. This time trace is well reproduced by the evolution of the excited-state population (red curve) under resonant excitation whose rising and falling edges match the characteristics of the experimental pulse (see Fig. S7). In particular, the slow rise in fluorescence before the front edge of the EOM is due to residual excitation of the molecule during the AOM rising edge. Nevertheless, our measurements demonstrate the ability of fast, sub-ns preparation of the molecular excited state with a probability as high as 0.97 if the duration of excitation is set to a half Rabi period.

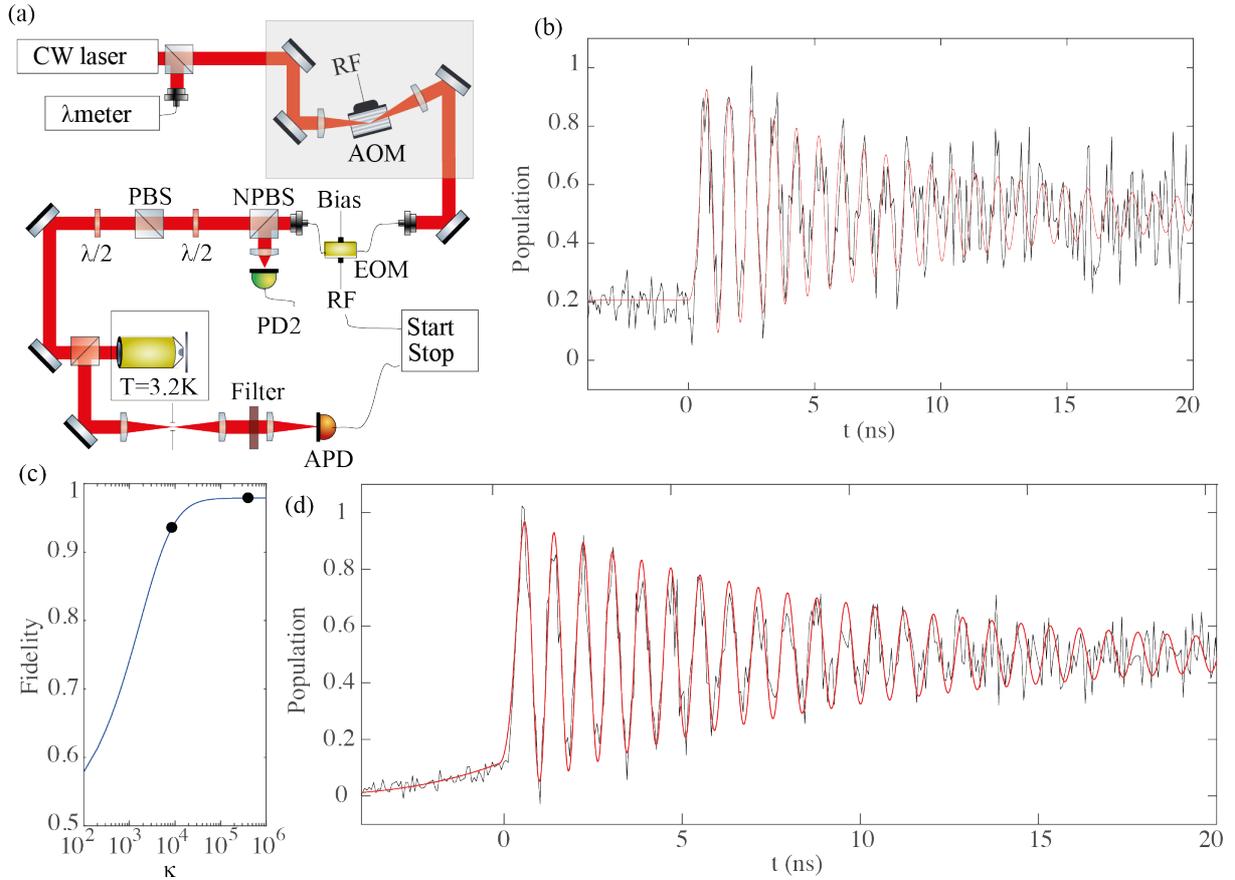

Figure 2 : Fast, sub-ns optical nutation of a single molecule. **a)** Optical setup designed to sculpt optical pulses in a single-frequency excitation laser beam, whose frequency is controlled with a wave meter to within 1 MHz. A confocal microscope placed inside a cryostat is used to excite and collect the fluorescence from a single DBATT molecule. The red-shifted fluorescence is spatially and spectrally filtered, using respectively a pinhole and a bandpass filter centered on 700 nm with a width of 100 nm, then sent onto an avalanche photodiode (APD). **b)** Temporal response of the fluorescence signal under resonant laser pulses shaped with the EOM alone with an intensity $I = 6100\, I_{sat}$. For such excitation intensity, there is no thermally induced decoherence, as demonstrated in the saturation plot of Fig. 1. The time trace is monitored by time-correlated single photon counting, and its origin of time corresponds to the rising edge of the EOM-shaped pulse. The red curve is the simulated evolution of the excited-state population, taking a rectangular shaped illumination profile, $T_2 = 2T_1$ and $\Omega_L T_1 = 55.2$. **c)** The theoretical fidelity is derived from the time evolution of the master equation calculated for a flat-top pulse for a given on-off intensity ratio $\kappa$ and corresponds to the maximum value reached by the population of the excited state. The black points are associated to the experimental configurations with the EOM alone ($\kappa \sim 8\times10^3$) and with both EOM and AOM ($\kappa \sim 3\times10^5$). With both EOM and AOM, we achieve the maximal fidelity obtainable with this excitation intensity. **d)** Optical nutation recorded under resonant excitation slots shaped with both the EOM and the AOM, at $I/I_{sat} = 6100$. The red curve is the simulated excited-state population, using a realistic temporal profile for the excitation (see Fig. S7). The fluorescence time traces **b, d** are scaled to the excited-state population, after background subtraction (Fig. S3, S4) and deconvolution from the response function of the avalanche photodiode (Fig. S5).

## IV. COHERENT MANIPULATION

Coherent manipulation of the molecular electronic state, such as population inversion or superposition of the ground and excited states with maximal coherence, requires controlled pulse areas $\mathcal{A} = \Omega_L \tau_P$, where $\tau_P$ is the duration of the flat-top excitation pulse and $\Omega_L$ is the Rabi frequency. With a view to fast coherent state preparation, we have submitted the molecule to a series of resonant pulses with fixed duration $\tau_P = 1$ ns and increasing excitation intensities, while collecting the fluorescence signal during the decay following each pulse, as sketched in Fig. 3a. Figures 3b-d exemplify fluorescence time traces of the molecule under pulsed excitation with $\mathcal{A} = \pi/2$, $\pi$ and $2\pi$ respectively, while the laser frequency is locked at resonance on the ZPL within 1 MHz. In the case of pulsed excitation with $\mathcal{A} = \pi/2$ (Fig. 3b), the population reaches ~0.5 after the pulse, which is consistent with the creation of a superposition between the molecular ground and excited states with maximal coherence. For pulsed excitation with $\mathcal{A} = \pi$ (Fig. 3c), the population of the molecule is inverted with a probability 0.97 at the end of the pulse. In the case of a $2\pi$ pulse (Fig. 3d), the molecular state is brought back to the ground state at the end of the pulse. The residual tail observed in the fluorescence temporal profile is explained by the finite slope of the AOM falling front (green profile in Fig. 3a), which causes residual excitation of the molecule at the end of the pulse. The evolution of the averaged fluorescence signal with increasing excitation intensities is shown in Fig. 3e (blue points). It is mainly characterized by oscillations, which reflect the tunability of the excited-state population after a controlled number of Rabi oscillations. Additionally, the heights of the minima increase with the pulse intensity due to the increased residual pumping of the molecule at the end of the pulse. Overall, the evolution of the population is reproduced with an exponentially decaying sinusoid with a linear offset (red curve in Fig. 3e), as in previous reports [24,31] but with a higher modulation depth.

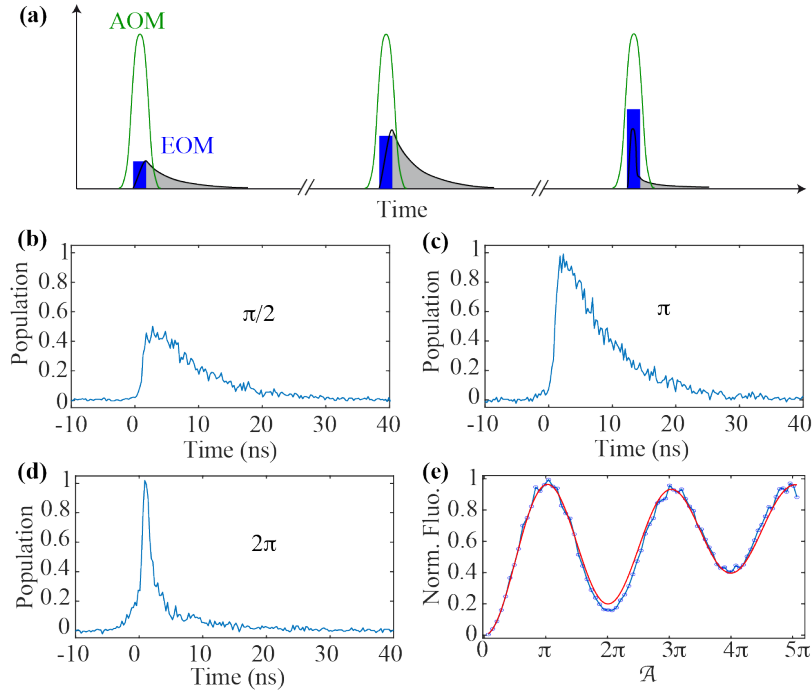

Figure 3: **a)** Sketch of the pulse sequence applied to the molecule for coherent state preparation. An AOM pulse (green profile) is superimposed on each EOM pulse (square-shaped intensity profiles in blue) to improve the on-off extinction ratio and reduce the stationary population of the excited state to less than 0.02 in the off periods. The black curves illustrate the temporal evolution of the excited-state population. The shaded light grey areas correspond to the periods (duration 7 ns) where the fluorescence signal is integrated. The pulses are separated by a 70 ns delay to let the system relax to its ground state. **b)-d)** Normalized time traces of the fluorescence signal under pulsed excitation in the cases $\mathcal{A} = \pi/2$ (b), $\mathcal{A} = \pi$ (c) and $\mathcal{A} = 2\pi$ (d). The fluorescence signal is scaled with that collected after a strong and long pulse producing a highly saturated population level (0.5) at its end. **e)** The normalized fluorescence signal, averaged over many pulse sequences, is displayed as a function of the pulse area. The data are fitted to an exponentially decaying sinusoid with a linear offset (red curve).

## V.  RAMSEY SPECTROSCOPY

In order to characterize the robustness of the molecular transition dipole against decoherence induced by electron-phonon interaction, we have investigated the temperature dependence of the dephasing rate using single-emitter Ramsey spectroscopy [7,28,29,32,33]. The evolution of the Bloch vector under the application of two pulses with duration $\tau_P$ and inter-pulse delay $\tau$ is illustrated in Fig. 4a, where $u$ and $v$ respectively correspond to the in-phase and in-quadrature responses of the dipole to the driving laser field, while $w$ is the population inversion between the excited and ground electronic states of the molecule. Between the two pulses, the Bloch vector undergoes a precession that is governed by the laser detuning $\delta_L$ with respect to the molecular resonance. As a consequence, the excited-state population obtained at the end of the second pulse oscillates with $\delta_L$. This generates Ramsey fringes of the fluorescence signal as the laser frequency is swept across the molecular resonance, as exemplified in Fig. 4b for $\tau_p = 1$ ns and $\tau = 2.7$ ns. In the case of infinitely short pulses, the fringe period is expected to evolve as $1/\tau$ [29]. Our calculations based on Bloch equations indicate that the Ramsey fringe periodicity should actually evolve as $1/(\tau + 4\tau_p/\pi)$, as demonstrated in Supplementary Note 1. This evolution is verified experimentally as the delay $\tau$ is increased (Figs. 4 b-d). Our numerical simulations of the excited-state population further support this analysis and reproduce well the experimental Ramsey fringes (red curves in Figs. 4 b-d).

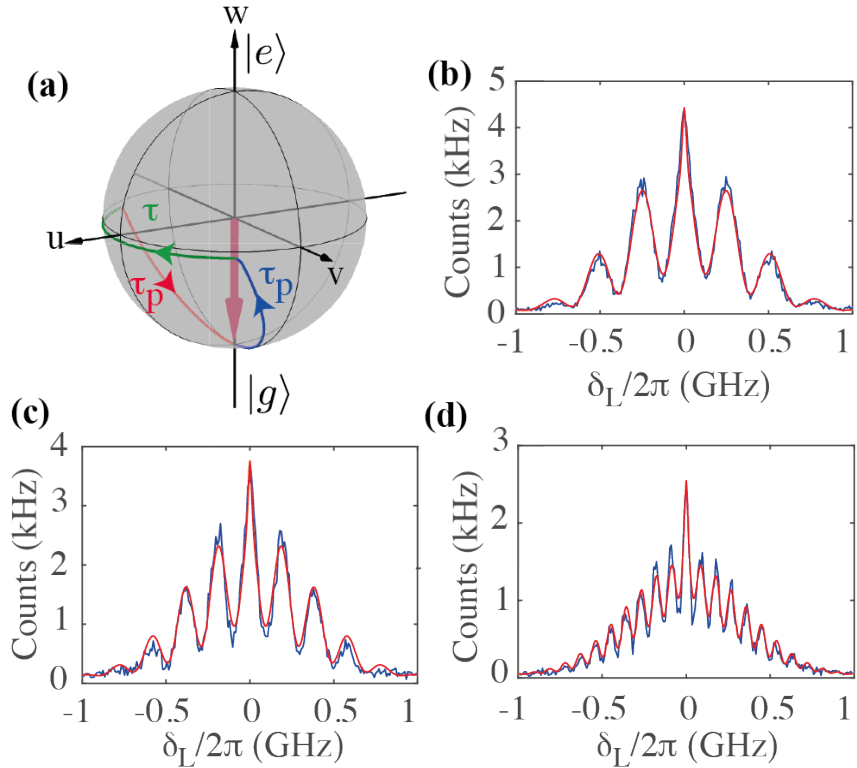

Figure 4: **a)** The top frame exemplifies the evolution of the Bloch vector of a two-level system initially set in the ground state (Bloch vector pointing downwards) and submitted to two consecutive pulses. The blue line represents the evolution of the Bloch vector during the first pulse; the green curve traces its free evolution between the pulses, while the red curve depicts its trajectory during the second pulse. At the end of this pulse sequence, the excited state population is sensitive to the pulse area, the laser detuning and the delay. **b)-d)** Ramsey fringes are obtained from the fluorescence signal when the laser is scanned through the ZPL transition, while series of pulse pairs with fixed intensity and duration ($\tau_p = 1$ ns) are applied, using various delays $\tau$: $\tau = 2.7$ ns in **b)**, $\tau = 4$ ns in **c)**, $\tau = 10$ ns in **d)**. The corresponding periods of the Ramsey fringes are 250 MHz, 187 MHz and 86 MHz, respectively, in accordance with the expected period given by $1/(\tau + 4\tau_p/\pi)$, considering a flat-top pulse.

While the derivation of $T_2$ from the ZPL width may be distorted by spectral diffusion, single-molecule Ramsey spectroscopy provides a direct determination of $T_2$. Indeed, the fluorescence signal measured after the second pulse decays with the delay $\tau$ between the two pulses, as a result of the decay of the coherence prepared by the first pulse, as illustrated in Fig. 5a. This decay time is exactly $T_2$ in the case of resonant $\pi/2$ pulses [29]. The upper inset of Fig. 5b shows the decay of the fluorescence signal as a function of $\tau$ at 3.2 K. The exponential fitting curve provides $T_2 = 14.7 \pm 0.7$ ns, which reaches its upper limit $2T_1$ and thus shows evidence for the absence of decoherence of the molecular transition dipole at this temperature. When raising the temperature to 9.8 K, $T_2$ drastically shortens to 1.6 ns, as evidenced in the lower inset of Fig. 5b. The temperature dependence of the dephasing rate $(\pi T_2)^{-1}$ measured with Ramsey spectroscopy increases exponentially with temperature, as shown in Fig. 5b (blue points). It is well reproduced with an Arrhenius law, with an activation energy $E_a = 42.9 \pm 0.8$ cm$^{-1}$ attributed to a quasi-local mode identified in the fluorescence spectrum [10,30]. This dephasing rate matches perfectly with the ZPL width (black points in Fig. 5b), which demonstrates the absence of spectral diffusion for this molecule.

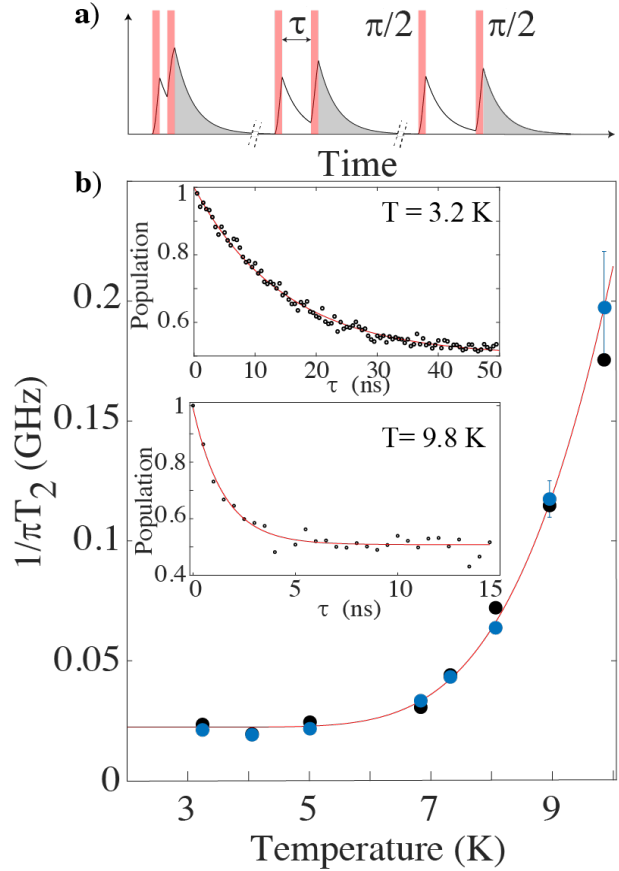

Figure 5: **a)** Scheme of the Ramsey pulse sequence (red bars), together with the evolution of the excited-state population (black curve) of a two-level system. The grey shaded areas after the second pulse indicate the periods over which the fluorescence signal is integrated to evaluate the excited state population. **b)** Evolution of the ZPL width (black points) and the dephasing rate $(\pi T_2)^{-1}$ measured from Ramsey spectroscopy (blue points), when temperature is increased. These evolutions match very well and are fitted with an Arrhenius law $(\pi T_2)^{-1}(T) = (\pi T_2)^{-1}(0) + A\, exp(-E_a/k_B T)$, with $A$ = 92.2 GHz and $E_a$ = 1287 GHz, while $(\pi T_2)^{-1}(0)$ is set to $\Delta \nu_0 = $ 22.3 MHz (red curve). The upper and lower insets respectively display the decay of the fluorescence signal at 3.2 K and 9.8 K as a function of the inter-pulse delay $\tau$. The exponential fits (red curves) provide a measure of $T_2$: 14.7 ± 0.7 ns at 3.2 K and 1.6 ± 0.2 ns at 9.8 K.

## VI. CONCLUSIONS

Interestingly, our Ramsey measurements of $T_2$ point to a vanishing dephasing rate at temperatures lower than 5 K, which are readily achievable with standard cryogen-free cooling technologies. This makes single molecules attractive electronic-state qubit systems that can be manipulated with near-unity fidelity and at rates much higher than the dephasing rate. Further studies will aim at exploring the coherence lifetime of entangled electronic states of dipole-dipole coherently coupled molecules [34,35]. In particular, the long-lived subradiant state that can be selectively excited [35] should open up extended coherence times for various quantum information schemes [36,37].

**Supplementary material :**

See the supplementary material (Figures S1 to S7) for experimental details and the supplementary note 1 for the derivation of Ramsey fringe periodicity.

**Acknowledgments :**

The authors acknowledge the financial support from the French National Agency for Research, Région Aquitaine (FAWN, 2018-1R50313), Institut Universitaire de France (B. Lounis), Idex Bordeaux (LAPHIA Program, Risky- MOmENTuM-AAP2020), the EUR Light S&T Graduate Program (PIA3 Program "Investment for the Future", ANR-17-EURE- 0027), and GPR LIGHT (LIGHT-006-IMOON-2021). We thank Eric Cormier for the loan of the arbitrary wave generator (AWG7122C).

**Author declaration:**

**Conflict of Interest**
The authors have no conflicts to disclose.

**Data Availability**

The data that support the findings of this study are available from the corresponding author upon reasonable request.